\documentclass[showpacs,preprintnumbers,amsmath,amssymb]{revtex4}
\usepackage{graphicx}
\usepackage{dcolumn}
\usepackage{bm}

\begin{document}

\title{Aspects of Noncommutative Scalar/Tensor Duality}
\author{Ajith K. M$^\ast$,~E. Harikumar$^\ast$,~Victor O. Rivelles$^\dagger$~and~M. Sivakumar$^\ast$}
\affiliation{
$^\ast$School of Physics, University of Hyderabad,\\ Hyderabad India
500046
\\{$^\dagger$}Instituto de F\'\i sica,
   Universidade de S\~ao Paulo \\ C. Postal 66318, 05315-970, S\~ao Paulo, SP,
  Brazil \\ ph01ph13,harisp,mssp@uohyd.ernet.in, rivelles@fma.if.usp.br}
\date{\today}
\begin{abstract}
We study the noncommutative massless Kalb-Ramond  gauge field
coupled to a dynamical  $U(1)$ gauge field in the adjoint
representation  together with a compensating vector field. We
derive the Seiberg-Witten map and obtain the corresponding mapped
action to first order in $\theta$. The
(emergent) gravity structure found in other situations is not present
  here. The 
off-shell dual scalar theory  is derived and it does not coincide
with the Seiberg-Witten mapped scalar theory. Dispersion relations
are also discussed. The p-form generalization of the
Seiberg-Witten map to order $\theta $ is also derived.
\end{abstract}

\maketitle

\section{Introduction}

Field theories, specially gauge theories in noncommutative (NC)
space-time have been actively studied in recent times. Maxwell and
Yang-Mills gauge theories have been extensively analyzed \cite
{rev}. There also exists  generalizations of massless vector
fields to p-form theories which are interesting by
themselves and also arise in several contexts in string theory,
field theory and condensed matter. It is well-known that in 4 
dimensions the 2-form theory, also known as the Kalb-Ramond (KR)
theory,  is dual to a massless scalar theory \cite{KR}. Hence it
is of interest to ask whether this feature remains true in a NC
space-time as well. Dual field  theories in NC space-time, using
the Seiberg-Witten (SW) map \cite{sw}, have been extensively
studied. These include the Maxwell-Chern-Simons and the self-dual
theories \cite{HV}, Bose-Fermi equivalence \cite{ahs}, etc. In all
these theories, the 
dual relation established in the commutative case does not carry
forward to the NC space-time. This work is devoted to the study of
the duality aspects of the abelian 2-form gauge theories in
non-commutative space-time.

The U(1) gauge theory in NC space-time has a non-abelian like
structure, and hence the abelian KR in NC space-time can also be
expected to have such a non-abelian like structure as well. The
naive generalization of the abelian  Lagrangian, viz $L = H*H$,
(where $H = dB$ and $B$ is the 2-form field) will not lead to any
NC correction as the star product can be removed from the
quadratic piece in the action. As it is well known, a Yang-Mills generalization of
the self-interacting KR theory does not exist and the closest non-abelian
generalization of the 2-form theory have been proposed in \cite {D,
FT}. In these models, KR fields transforming in the adjoint
representation of the gauge group couple with non-abelian gauge
fields.



We consider a massless KR field transforming in the adjoint
representation of $U(1)$ (a possibility which does not exist in
commutative space-time) coupled with the gauge field. The study of
such a theory has also another motivation. The existence of
several parallels between NC U(1) gauge theory and commutative
gravitational theory, led one of us to seek a direct relationship
between them. It was observed that a massless scalar and vector
fields coupled to NC $U(1)$ gauge field, after the SW mapping, has
the same structure as that of fields coupled with gravity
\cite{VOR}.  This suggestion, that gravity can emerge from NC
electromagnetism,  has been further studied and extended in
\cite{yang, sten}. Hence it is natural to ask whether the 2-form
field coupled to photons in NC space-time has such an
emergent gravity coupling. First we discuss the NC KR theory. We
recall the construction for the non-abelian case, where a
topological coupling to a vector field with mass parameter was
also considered \cite {lahiri, hwang} and show its extension to
the NC case in section 2. Then, in section 3, we discuss the SW
map and the emergent gravity picture. Section 4 is devoted to the
analysis of the duality between both NC formulations while in
section 5 we discuss the dispersion relations. Finally, in section
6 we present our conclusions and discuss the extension of our
results to the case of a p-form field.

\section{NC Kalb-Ramond theory}
The NC action for a massless Kalb-Ramond 2-form field is given
by
  \cite{lahiri,hwang}

\begin{equation}
\int{d^{4}x ~~(\frac{1}{12}{\hat{\cal
H}}_{\mu\nu\lambda}\star{\hat{\cal
H}}^{\mu\nu\lambda}}-\frac{1}{4}\hat F_{\mu\nu}\star\hat
F^{\mu\nu}),
 \label{nkr}
\end{equation}
where
\begin{eqnarray}
{\hat{\cal H}}_{\mu\nu\lambda}&=&D_{\mu}\hat {{\cal
B}}_{\nu\lambda}+ {\rm cyclic~~ terms}, \nonumber\\
&=&D_{\mu}[\hat B_{\nu\lambda}-(D_{\nu}\hat
C_{\lambda}-D_{\lambda}\hat C_{\nu})]+ {\rm cyclic~~ terms}, \\
\hat F^{\mu\nu} &=& -i[D^{\mu},D^{\nu}]_{\star} =
-i(\partial_{\mu}A_{\nu}-\partial_{\nu}A_{\mu}-i[A_{\mu},A_{\nu}]_{\star}),
\end{eqnarray}
and the action of the covariant derivative is defined by
\begin{equation}
D_{\mu}\hat B_{\nu\lambda}=\partial_{\mu}\hat
B_{\nu\lambda}-i[\hat A_{\mu},\hat B_{\nu\lambda}]_{\star}.
\end{equation}

 The field equation for $B^{\nu\lambda}, C_{\lambda}$
and $A_{\nu}$ are given, respectively, by

\begin{equation}
D^{\mu}{\hat{\cal H}}_{\mu\nu\lambda}=0,
 \label{1neqm}
\end{equation}
\begin{equation}
[{\hat{\cal H}}_{\mu\nu\lambda}, \hat F^{\mu\nu}]_*=0,
 \label{2neqm}
 \end{equation}
 \begin{equation}
D^{\mu}\hat F_{\mu\nu} +  2i[{\hat C}^{\lambda},D^{\mu}{\hat{\cal
H}}_{\mu\nu\lambda}]_{\star}-i[{{\cal B}}^{\mu\lambda},{\hat{\cal
H}}_{\mu\nu\lambda}]_{\star}=0. \label{3neqm}
\end{equation}

The generalization of the commutative  KR theory to NC space-time
requires the Maxwell term, without which the
theory becomes trivial. This can be easily seen from the fact that
the first term in Eqn.(\ref {3neqm}) will be absent when the
$F^{\mu\nu}F_{\mu\nu}$ is not present in the Lagrangian. In that
case, substituting Eqn.(\ref{1neqm}) into Eqn.(\ref{3neqm}) will
result in that either ${\cal B}$ or ${\cal H}$ is a constant
leading to the absence of any effect of noncommutativity in
Eqn.(\ref{nkr}). Also, the need for the C field  becomes clear
from the consistency of the field equations given above: the
divergence of Eqn.(\ref{1neqm}) will vanish only if we use the
equation Eqn.(\ref{2neqm}). 

The local symmetries of Eqn(\ref{nkr})
are:

1) NC U(1) gauge transformations :
\begin{eqnarray}
\label{gtv}
\delta \hat B_{\mu\nu}&=& i[\hat \lambda,\hat
B_{\mu\nu}]_{\star}, \nonumber\\
 \delta \hat C_{\mu}& =&i[\hat\lambda, \hat C_{\mu}]_{\star}, \nonumber\\
\delta \hat A_{\mu}&=& D_{\mu}\hat \lambda,
\end{eqnarray}

2) the local Kalb-Ramond transformations, with a vector
parameter, given by
\begin{eqnarray}
\label{gts}
\delta \hat B_{\mu\nu}&=&D_{\mu}\hat
\Lambda_{\nu}-D_{\nu}\hat
\Lambda_{\mu}, \nonumber\\
\delta \hat C_{\mu}&=&\hat \Lambda_{\mu}, \nonumber\\
\delta \hat A_{\mu}&=&0.
\end{eqnarray}
{This defines the NC U(1) KR theory.}

\section{The SW map for the NC U(1) KR theory}
In this section we derive the SW map for NC U(1) KR theory.
Earlier the SW map for the non-abelian 2-form theories has been studied by
using the closure of the algebra\cite{amo}. Here we derive the mapping by solving
explicitly the SW equation.To
derive the SW map we consider the complete gauge transformations
of  $\hat B_{\mu\nu}$,
$\hat C_{\mu}$ and $ \hat A _{\mu}$

\begin{eqnarray}
\delta \hat A_{\mu}&=&D_{\mu}\hat \lambda, \nonumber\\
\delta \hat B_{\mu\nu}&=&i[\hat\lambda,\hat
B_{\mu\nu}]_{\star}+(D_{\mu}\hat
\Lambda_{\nu}-D_{\nu}\hat\Lambda_{\mu}), \nonumber\\
\delta \hat C_{\mu}&=&i[\hat \lambda,\hat C_{\mu}]_{\star}+\hat
\Lambda_{\mu}.
\end{eqnarray}

The SW map is obtained by demanding that the gauge orbit of the NC
gauge field gets mapped to that of commutative theory.
In the case of the vector field, under the U(1) transformation,
the SW map is well known \cite{sw} and the SW equation for
$A_{\mu}$ is
\begin{equation}
\hat A(A, \theta) + \delta_{\hat\lambda} \hat A(A,\theta) = \hat
A(A + \delta_ {\lambda} A, \theta). \label{swa}
\end{equation}
To first order in $\theta$ we write $\hat A=A+A^{\prime}(A)$ where
$A^{\prime}$ is a function of A and  $\theta$. And for the gauge
parameter we write $\hat\lambda(\lambda,
A)=\lambda+\lambda^{\prime}(\lambda, A)$. Then Eqn.(\ref{swa})
becomes
\begin{equation}
{A}^{\prime}_{\mu}(A+\delta
A)-{A}^{\prime}_{\mu}(A)=\partial_{\mu}\lambda^{\prime}+\theta^{\rho\sigma}\partial_{\rho}A_{\mu}\partial_{\sigma}\lambda.
\label{swas}
\end{equation}
The solution for $A^{\prime}$ and $\lambda^{\prime}$ satisfying
the above equation is given by
\begin{eqnarray}
A^{\prime}&=&-\frac{1}{2}\theta^{\alpha\beta}A_{\alpha}(\partial_{\beta}A_{\mu}+F_{\beta\mu}),\nonumber\\
\lambda^{\prime}&=&\frac{1}{2}\theta^{\alpha\beta}\partial_{\alpha}\lambda
A_{\beta}.
\end{eqnarray}
Since the U(1) gauge field is inert under the KR
transformations the above equations retain their familiar form.

We now set up the SW equations for the fields $B_{\mu\nu}$ and
$C_{\mu}$ much as in the same way as given by Eqn.(\ref{swas})
\begin{eqnarray}
{B}^{\prime}_{\mu\nu}(B+\delta B,A+\delta A, C+\delta
C)-{B}^{\prime}_{\mu\nu}(B,A,C)&=&(\partial_{\mu}\Lambda^{\prime}_{\nu}-\partial_{\nu}\Lambda^{\prime}_{\mu})-\theta^{\rho\sigma}\partial_{\rho}\lambda\partial_{\sigma}B_{\mu\nu},\nonumber\\&+&
\theta^{\rho\sigma}[\partial_{\rho}A_{\mu}\partial_{\sigma}\Lambda_{\nu}-\partial_{\rho}A_{\nu}\partial_{\sigma}\Lambda_{\mu}
]\nonumber\\
{C}^{\prime}_{\mu}(C+\delta
C)-{C}^{\prime}_{\mu}(C)&=&\Lambda^{\prime}_{\mu}-\theta^{\rho\sigma}\partial_{\rho}\lambda\partial_{\sigma}C_{\mu}.
\label{swe}
\end{eqnarray}
The solution of the SW equations to first order in $\theta$ is
\begin {eqnarray}
\hat B_{\mu\nu}&=&B_{\mu\nu}+B^{\prime}_{\mu\nu} \nonumber\\
&=&B_{\mu\nu}-\theta^{\rho\sigma}(A_{\rho}\partial_{\sigma}B_{\mu\nu}+\partial_{\mu}A_{\rho}\partial_{\sigma}C_{\nu}-\partial_{\nu}A_{\rho}\partial_{\sigma}C_{\mu}\nonumber\\
&-&(\partial_{\rho}A_{\mu}\partial_{\sigma}C_{\nu}-\partial_{\rho}A_{\nu}\partial_{\sigma}C_{\mu})),\nonumber\\
\hat\Lambda_{\mu}&=&\Lambda_{\mu}+\Lambda^{\prime}_{\mu}\nonumber\\
&=&\Lambda_{\mu}-\theta^{\rho\sigma}A_{\rho}\partial_{\sigma}\Lambda_{\mu},\nonumber\\
\hat C_{\mu}&=&C_{\mu}+C^{\prime}_{\mu}\nonumber\\
&=&C_{\mu}-\theta^{\rho\sigma}A_{\rho}\partial_{\sigma}C_{\mu}.
\label{sws1}
\end{eqnarray}
It is easy to verify the closure of the algebra
\begin{eqnarray}
[\delta_{\Lambda_{1}},\delta_{\Lambda_{2}}]B^{\prime}_{\mu\nu}=0,~
[\delta_{\Lambda},\delta_{\lambda}]B^{\prime}_{\mu\nu}=0,~
[\delta_{\Lambda_{1}},\delta_{\Lambda_{2}}]C^{\prime}_{\mu}=0,~
[\delta_{\Lambda},\delta_{\lambda}]C^{\prime}_{\mu}=0
\end{eqnarray}

The next step is to derive the SW mapped action using the solution
above. The SW mapped  field strength ${\hat{\cal
H}}_{\mu\nu\lambda}$ is given to first order in $\theta$ by
\begin{eqnarray}
\label{fssw} {\hat{\cal
H}}_{\mu\nu\lambda}=H_{\mu\nu\lambda}(B)+\theta^{\rho\sigma}(F_{\rho\mu}\partial_{\sigma}{\cal
B}_{\nu\lambda}+\hbox{Cyclic
terms})-\theta^{\rho\sigma}A_{\rho}\partial_{\sigma}H_{\mu\nu\lambda}
\end{eqnarray}
The gauge invariance of ${\hat{\cal H}}_{\mu\nu\lambda }$
under KR transformation and covariance under U(1) transformations
is evident.

Using Eqn.(\ref{fssw}) in Eqn.(\ref{nkr}) the SW
mapped action to first order in $\theta$ is
\begin{eqnarray}
S_{sw}&=&\int d^{4}x
(\frac{1}{12}H_{\mu\nu\lambda}H^{\mu\nu\lambda}
+\frac{1}{2}\theta^{\rho\sigma}H_{\mu\nu\lambda}F_{\rho}^{\mu}\partial_{\sigma}
[B^{\nu\lambda}-\partial_{\nu}C_{\lambda}+\partial_{\lambda}C_{\nu}]\nonumber\\&+&\frac{1}{24}\theta^{\rho\sigma}H_{\mu\nu\lambda}F_{\sigma\rho}H^{\mu\nu\lambda}\nonumber\\
&-&\frac{1}{4}[F_{\mu\nu}F^{\mu\nu}+2\theta^{\rho\sigma}F_{\rho\mu}F_{\sigma\nu}F^{\mu\nu}-\frac{1}{2}\theta^{\rho\sigma}F_{\rho\sigma}F^{\mu\nu}F_{\mu\nu}]).
\label{swa1}
\end{eqnarray}
The above action is invariant under the usual commutative gauge
transformations
\begin{eqnarray}
\delta B_{\mu\nu}=\partial_{[\mu}\Lambda_{\nu]},~\delta
A_{\mu}=\partial_{\nu}\lambda ~\hbox{and}~C_{\mu}=\Lambda_{\mu}.
\end{eqnarray}

The presence of C field in the SW mapped action is essential to
maintain the B field symmetry $\partial_{[\mu}\Lambda_{\nu]}$.
When we take the commutative limit $\theta\rightarrow0$, the field
C will disappear as it happens in the commutative case. Next we
must check the consistency of the field equations for $A_{\mu}$,
$B_{\mu\nu}$ and $C_{\lambda}$ derived from the action above. The
field equation for the field $A_\mu$ is
\begin{equation}
\partial_{\mu}F^{\mu\nu}+\theta^{\alpha\beta}F_{\alpha}^{~\mu}(\partial_{\beta}F_{\mu}^{~\nu}+
\partial_{\mu}F_{\beta}^{~\nu})-\frac{1}{2}\theta^{\alpha\beta}\partial_{\alpha}H_{\nu\tau\lambda}\partial_{\beta}{\cal
B}^{\tau\lambda}=0,  \label{aeq}
\end{equation}
while the equation for $B_{\nu\lambda}$ is
\begin{eqnarray}
\partial^{\mu}H_{\mu\nu\lambda}+\theta^{\rho\sigma}\partial^{\mu}(F_{\rho\mu}\partial_{\sigma}{\cal
B}_{\nu\lambda}+F_{\rho\nu}\partial_{\sigma}{\cal
B}_{\lambda\mu}+F_{\rho\lambda}\partial_{\sigma}{\cal
B}_{\mu\nu})+\theta^{\rho\sigma}\partial_{\sigma}H_{\mu\nu\lambda}F_{\rho}^{~\mu}=0,
\label{Beq}
\end{eqnarray}
and the $C_{\lambda}$ field equation  is
 \begin{equation}
 \theta^{\rho\sigma}\partial_{\rho}F^{\mu\nu}\partial_{\sigma}H_{\mu\nu\lambda}=[F_{\mu\nu},H_{\mu\nu\lambda}]_{\ast}=0.
 \label{ceq}
 \end{equation}
Notice that field equation for $C_{\lambda}$ is needed for the
consistency of the $B_{\mu\nu}$ field equation. The divergence of the B
equation will vanish only  if we use the C equation of motion. The
divergence of  both, the C and A  field equations are  zero
identically.

 It was shown by one of us \cite{VOR} that, after the
SW map, the coupling of a massless scalar field to a gauge field
in a NC space-time has the same structure as a gravitational
coupling. Hence it is of interest to inquire whether a similar
phenomenon takes place here too. In order to do that we note that
the SW mapped Lagrangian can be  rewritten as
\begin{eqnarray}
S&=&\int
d^{4}x\{\frac{1}{12}H_{\mu\nu\lambda}H^{\mu\nu\lambda}+[(\frac{1}{2}\theta^{\rho\sigma}F_{\rho}^{~\mu}-\frac{1}{8}\theta^{\alpha\beta}F_{\alpha\beta}\eta_{\mu\sigma})]H_{\mu\nu\lambda}\partial^{\sigma}{\cal
B}^{\nu\lambda}\}. \label{SWA}
\end{eqnarray}
Notice that at this point we are working with ${\cal B_{\mu\nu}} =
B_{\mu\nu} -\partial_{[\mu} C_{\nu]}$ and $ H (B ) = H ({\cal B})
$. Next we consider the coupling of the Kalb-Ramond field to
gravity
\begin{equation}
S_{g,B}=\int{d^{4}x
  \sqrt{-g}g^{\mu\alpha}g^{\nu\beta}g^{\lambda\sigma}H_{\mu\nu\lambda}H_{\alpha\beta\sigma}},
\end{equation}
and  the expansion of the metric $g_{\mu\nu}$ around flat
space-time $\eta_{\mu\nu}$ as
\begin{equation}
g_{\mu\nu}=\eta_{\mu\nu}+h_{\mu\nu}+\eta_{\mu\nu}h,
\end{equation}
where $h_{\mu\nu}$ is traceless. We then get
\begin{equation}
S_{g,B}=\int{d^{4}x(\frac{1}{12}H_{\mu\nu\lambda}H^{\mu\nu\lambda}-\frac{1}{12}hH_{\mu\nu\lambda}H^{\mu\nu\lambda}-\frac{1}{4}h^{\mu\alpha}H_{\mu\nu\lambda}[\partial_{\alpha}
{\cal B}^{\nu\lambda}+2\partial^{\nu}{\cal B}_{\lambda\alpha}])}.
 \label{SB}
\end{equation}
To find an expression for $h^{\alpha\mu}$ in terms of the gauge
fields we should compare Eqn.(\ref{SWA}) and Eqn.({\ref{SB}). But
the presence of a antisymmetric part for the term in the square
bracket in Eqn.(\ref{SWA}) destroys this comparison. Thus the NC
SW mapped KR theory does not have the structure of emergent
gravity as for the NC scalar field. \cite{VOR}.

\section{Duality between the KR and scalar fields}
It is well known that in four dimensions the massless KR theory is
dual to a massless scalar theory \cite {KR}. In this section we
will search for the dual field theory of the SW mapped KR theory.
We will apply the method developed in \cite{ESF}, which is an
off-shell dual formulation. We recast the action using an
auxiliary field $ A_{\alpha\beta\gamma} =\partial_{
\alpha}B_{\beta\gamma}$ enforced through a Lagrange multiplier $
\phi$. To implement the duality transformation we start with the
first order Lagrangian
\begin{eqnarray}
S_{sw}&=&\int d^{4}x~ (\frac{1}{4}(A_{\mu\nu\lambda}+2A_{\lambda\mu\nu})A^{\mu\nu\lambda}\nonumber\\
&+&\frac{1}{2}\theta^{\rho\sigma}(A_{\mu\nu\lambda}+2A_{\lambda\mu\nu})F_{\rho\mu}(A^{\sigma\nu\lambda}-\partial_{\sigma}(\partial^{\nu}C^{\lambda}-\partial^{\lambda}C^{\nu}))\nonumber\\
&+&\frac{1}{8}\theta^{\rho\sigma}(A_{\mu\nu\lambda}+2A_{\lambda\mu\nu})A^{\mu\nu\lambda}F_{\sigma\rho}\nonumber\\&+&\frac{1}{3}\phi^{\mu\nu\lambda}(A_{\mu\nu\lambda}-\partial_{\mu}B_{\nu\lambda}))+S^{sw}_{Maxwell}.
\label{fr}
\end{eqnarray}

Eliminating $\phi$ by using its field equation gives back the
original theory. If instead  we eliminate the original field B we
find a constraint on $\phi$, i.e,
$\partial_{\mu}\phi^{\mu\nu\lambda}=0$, which can be solved as
$\phi^{\mu\nu\lambda}=\epsilon^{\mu\nu\lambda\gamma}\partial_{\gamma}\Phi$.
The field equation for A is then given by
\begin{eqnarray}
&&A_{[\alpha,\beta\gamma]}(1+\frac{1}{2}\theta^{\rho\sigma}F_{\sigma\rho})+\theta^{\rho\sigma}(F_{\rho\alpha}A_{\sigma,\beta\gamma}+F_{\rho\gamma}A_{\sigma,\alpha\beta}+F_{\rho\beta}A_{\sigma,\gamma\alpha})\nonumber\\&+&
\theta^{\rho\alpha}F_{\rho}^{\mu}(A_{\mu,\beta\gamma}+A_{\gamma,\mu\beta}+A_{\beta,\gamma\mu})+\theta^{\rho\sigma}[F_{\rho\alpha}\partial_{\sigma}(\partial_{[\gamma}C_{\beta]})\nonumber\\&+&F_{\rho\beta}\partial_{\sigma}(\partial_{[\alpha}C_{\gamma]})+F_{\rho\gamma}\partial_{\sigma}(\partial_{[\beta}C_{\alpha]})]+\frac{1}{3}\epsilon_{\alpha\beta\gamma\tau}\partial^{\tau}\Phi=0.
\end{eqnarray}
Now we can solve it perturbatively in $\theta$ to get
\begin{eqnarray}
A_{[\alpha\beta\gamma]}&=&-\frac{1}{3}\epsilon_{\alpha\beta\gamma\lambda}\partial^{\lambda}\Phi+
\theta^{\rho\sigma}\frac{1}{(3)}
[F_{\rho\alpha}\epsilon_{\sigma\beta\gamma\lambda}\partial^{\lambda}\Phi+
F_{\rho\gamma}\epsilon_{\sigma\alpha\beta\lambda}\partial^{\lambda}\Phi+
F_{\rho\beta}\epsilon_{\sigma\gamma\alpha\lambda}\partial^{\lambda}\Phi]\nonumber\\
&+&\frac{1}{9}[\theta^{\rho}_{~\alpha}F_{\rho}^{~\mu}\epsilon_{\mu\beta\gamma\lambda}\partial^{\lambda}\Phi+(\alpha\rightarrow\beta\rightarrow\gamma)]+\theta^{\rho\sigma}\frac{1}{3!}\epsilon_{\alpha\beta\gamma\lambda}\partial^{\lambda}\Phi
~F_{\sigma\rho}\nonumber\\
&+&\theta^{\rho\sigma}[F_{\rho\alpha}\partial_{\sigma}(\partial_{\beta}C_{\gamma}-\partial_{\gamma}C_{\beta})+F_{\rho\beta}\partial_{\sigma}(\partial_{\gamma}C_{\alpha}-\partial_{\alpha}C_{\gamma})+F_{\rho\gamma}\partial_{\sigma}(\partial_{\alpha}C_{\beta}-\partial_{\beta}C_{\alpha})].\nonumber\\
\end{eqnarray}
Using this we can eliminate the field $A_{\mu\nu\lambda}$ from the
action Eqn.(\ref{fr}) to get the  dual Lagrangian
\begin{eqnarray}
L_{D}&=&\frac{1}{2}\partial_{\lambda}\Phi\partial^{\lambda}\Phi+\frac{5}{36}\theta^{\rho\sigma}F_{\rho\sigma}\partial_{\lambda}\Phi\partial^{\lambda}\Phi
-\frac{1}{6}\theta^{\rho\tau}F_{\rho\lambda}\partial_{\tau}\Phi\partial^{\lambda}\Phi\nonumber\\
&+&\frac{1}{3}\theta^{\rho\sigma}F_{\rho}^{~\alpha}\partial_{\sigma}(\partial^{\beta}C^{\gamma}-\partial^{\gamma}C^{\beta})\epsilon_{\alpha\beta\gamma\tau}\partial^{\tau}\Phi+\hbox{Maxwell's
  term}.
\label{dlag}
\end{eqnarray}

Now from this dual Lagrangian the field equations for $\Phi$, A
and C to order $\theta $ are given by
\begin{equation}
-\partial_{\lambda}\partial^{\lambda}\Phi-\frac{7}{36}\theta^{\alpha\beta}\partial_{\lambda}F_{\alpha\beta}\partial^{\lambda}\Phi
+\frac{1}{3}\theta^{\rho\tau}F_{\rho\lambda}\partial^{\lambda}\partial_{\tau}\Phi+\frac{2}{3}\theta^{\rho\sigma}\epsilon_{\alpha\beta\gamma\tau}\partial^{\tau}F_{\rho}^{~\alpha}\partial_{\sigma}\partial^{\beta}C^{\gamma}=0,
\end{equation}
\begin{eqnarray}
&&\partial_{\mu}F^{\mu\sigma}+\theta^{\alpha\beta}F_{\alpha}^{~\mu}(\partial_{\beta}F_{\mu}^{~\sigma}+
\partial_{\mu}F_{\beta}^{~\sigma})-\frac{5}{18}\theta^{\rho\sigma}\partial_{\rho}(\partial_{\lambda}\Phi\partial^{\lambda}\Phi)+\frac{1}{6}\theta^{\rho\tau}\partial_{\rho}(\partial_{\tau}\phi\partial^{\sigma}\Phi)\nonumber\\&-&\frac{1}{6}\theta^{\sigma\tau}\partial_{\lambda}(\partial_{\tau}\Phi\partial^{\lambda}\Phi)
+\frac{2}{3}\theta^{\rho\alpha}\partial_{\alpha}\partial_{\beta}C_{\gamma}\epsilon_{\sigma\beta\gamma\tau}\partial_{\rho}\partial^{\tau}\Phi
=0, \label{aeq1}
\end{eqnarray}
\begin{equation}
[F_{\mu\nu},\partial_{\sigma}\tilde\Phi]_{\ast}=0.
\label{ceq1}
\end{equation}
The C field equation on both sides of the dual theory (i.e.,
Eqn.(\ref{ceq}) and Eqn.(\ref{ceq1})) is consistent with the
familiar dual relation
$\partial_{\mu}\phi=\frac{1}{3!}\epsilon_{\mu\nu\lambda\sigma}H^{\nu\lambda\sigma}$

In \cite{VOR} the SW mapped scalar theory is given by
\begin{eqnarray}
\label{action_scalar_field} S_\varphi = \frac{1}{2} \int d^4x \,
\left[ \partial^\mu \varphi
\partial_\mu \varphi + 2 \theta^{\mu\alpha} {F_\alpha}^\nu \left(
-
\partial_\mu \varphi
\partial_\nu \varphi + \frac{1}{4} \eta_{\mu\nu} \partial^\rho \varphi
\partial_\rho \varphi \right) \right].
\end{eqnarray}
Notice that the structure of the dual theory Eqn.(\ref{dlag}) is
different from the above one and also that it does not have the
form of a  gravity coupling.
\section{Dispersion relation}
In this section we will look for the solutions to the field
equations in order to derive the dispersion relations. Dispersion
relations in Maxwell equations case have been well studied
\cite{disp}. The field equations are given by (\ref{aeq}) and
(\ref{Beq}).
 There is a trivial solution where F and ${\cal B}$ are both
constants. If ${\cal B}$ is constant then the dispersion relation
for F remains the same as that of the SW mapped Maxwell case
alone. Next, we look for a transverse plane wave solution for
$A_{\mu}$ and ${\cal B}_{\mu\nu}$.  To this end we define
$A_{\mu}=\tilde{A}_{\mu}(kx)$ and  ${\cal
B}_{\mu\nu}$=${\tilde{\cal
  B}}_{\mu\nu}(kx)$, with $k$ being a wave vector.
Using $\partial_{\alpha}A_{\mu}=k_\alpha \tilde A_{\mu}^{\prime}$
where the prime denotes differentiation w.r.t to $kx$. Then the
field equation for $A_\mu$ becomes
\begin{eqnarray}
&&k^{2}\tilde A_{\nu}^{\prime}-k_{\mu}k_{\nu}\tilde
A^{\prime\mu}+\theta^{\alpha\beta}(k_{\alpha}\tilde
A^{\prime\mu}-k^{\mu}\tilde
A_{\alpha}^{\prime})[k_{\beta}k_{\mu}\tilde
A_{\nu}^{\prime}-k_{\beta}k_{\nu}\tilde
A_{\mu}^{\prime}+k_{\mu}k_{\beta}\tilde
A_{\nu}^{\prime}-k_{\mu}k_{\nu}\tilde
A_{\beta}^{\prime}]\nonumber\\&-&\frac{\theta^{\alpha\beta}}{2}k_{\alpha}(k_{\nu}\tilde{\cal
  B}_{\tau\lambda}^{\prime}+k_{\lambda}\tilde {\cal
  B}_{\nu\tau}^{\prime}+k_{\tau}\tilde {\cal
  B}_{\lambda\nu}^{\prime})k_{\beta}\tilde {\cal B}^{\prime
  \tau\lambda}=0.
\end{eqnarray}
This reduces to
\begin{equation}
k^{2}(1-2\theta^{\alpha\beta}k_{\alpha}\tilde
A_{\beta}^{\prime})\tilde A_{\nu}^{\prime}=0
\end{equation}
by use of the transversality conditions $k_{\alpha}\tilde
A^{\alpha}=k_{\alpha}\tilde{\cal B}^{\alpha\beta}=0$. This
produces the usual dispersion relation $k^{2}=0$. Now we have to
check the consistency with B field equation. It is given by

\begin{eqnarray}
&&k^{\mu}(k_{\mu}\tilde{\cal B}_{\nu\lambda}^{\prime}
+k_{\lambda}\tilde{\cal B}_{\mu\nu}^{\prime}+k_{\nu}\tilde{\cal B}_{\lambda\mu}^{\prime})
+\theta^{\rho\sigma}k^{\mu}[(k_{\rho}A_{\mu}-k_{\mu}A_{\rho})k_{\sigma}\tilde{\cal B}^{\prime}_{\nu\lambda}\nonumber\\
&+&(k_{\rho}A_{\nu}-k_{\nu}A_{\rho})k_{\sigma}\tilde{\cal B}^{\prime}_{\mu\lambda})
+(k_{\rho}A_{\lambda}-k_{\lambda}A_{\rho})k_{\sigma}\tilde{\cal B}^{\prime }_{\mu\nu}]\nonumber\\
&+&\theta^{\rho\sigma}k_{\sigma}(k_{\mu}\tilde{\cal
B}^{\prime}_{\nu\lambda}+k_{\lambda}\tilde{\cal
B}_{\mu\nu}^{\prime}+k_{\nu}\tilde{\cal
B}_{\lambda\mu}^{\prime})(k_{\rho}\tilde
A^{\prime\mu}-k^{\mu}\tilde A_{\rho}^{\prime})=0.
\end{eqnarray}

The transversality condition of B will reduce this further to
\begin{equation}
k^{2}(1-2\theta^{\alpha\beta}k_{\alpha}\tilde
A_{\beta}^{\prime}){\cal B}^{\prime}_{\nu\lambda}=0,
\end{equation}
so that we have the usual dispersion relation $k^{2}=0$. Next we
look for a solution where
 F is  constant and H is a plane wave
 From Eqn.(\ref{Beq}) and using the transversality
condition for $\cal{B}$, we get
\begin{eqnarray}
\tilde{k}_\lambda H^{\prime\lambda\mu\nu} + \theta^{\alpha\beta}
F_{\alpha\lambda} k_\beta k^\lambda{\cal B}^{\prime\mu\nu}=0,
\end{eqnarray}
where we have defined $\tilde{k}_{\mu}=k_{\mu} +
\theta^{\alpha\beta} F_{\alpha\mu} k_{\beta}$.
 The Bianchi identity (BI) for H is given by
\begin{equation}
k_{\mu} \tilde H_{\nu\lambda\rho}^{\prime}-k_{\rho}\tilde
H_{\mu\nu\lambda}^{\prime}+k_{\lambda} \tilde
H_{\rho\mu\nu}^{\prime}-k_{\nu} \tilde H_{\lambda\rho\mu}^{\prime}
= 0.
\end{equation}
Now we contract the BI with $\tilde{k}^\mu$. The $\tilde {\cal
  B}^{\prime}$ terms can
be collected together as $\tilde H^{\prime}$ and we get
\begin{eqnarray}
( k \tilde{k} + \theta^{\alpha\beta} F_{\alpha\sigma} k_\beta
k^\sigma ) \tilde H_{\nu\lambda\rho}^{\prime} = 0,
\end{eqnarray}
so the dispersion relation is
\begin{equation}
k^2 = - 2 \theta^{\alpha\beta} F_{\alpha\lambda} k_\beta k^\lambda.
\end{equation}
Now adopting the same conventions as in \cite{VOR} for the
decomposition into transversal and longitudinal components this
will lead to
\begin{equation}
\frac{\vec k^{2}}{\omega^{2}}=1-2\vec \theta_{T} \cdot (\vec
B_{T}-\frac{\vec k}{\omega}\times \vec E_{T}).
\end{equation}
A similar relation was obtained in \cite{VOR} for the scalar field.
Though the dual equivalence between KR and
scalar fields at the level of the partition function is absent, at
the level of the dispersion relations they agree.
\section{Conclusion}
In this work we have studied the scalar/tensor duality in NC
space-time. The SW equation was set up and solved. The solution
given in (\ref{sws1}) is not unique. There exists one more
solution below:
\begin{eqnarray}
\hat B_{\mu\nu}&=&B_{\mu\nu}+B^{\prime\prime}_{\mu\nu}\nonumber\\
&=&B_{\mu\nu}-\theta^{\rho\sigma}(\partial_{\mu}A_{\rho}B_{\sigma\nu}
-\partial_{\nu}A_{\rho}B_{\sigma\mu}+A_{\rho}\partial_{\sigma}B_{\mu\nu}\nonumber\\
&-&\partial_{\rho}A_{\mu}B_{\sigma\nu}+\partial_{\rho}A_{\nu}B_{\sigma\mu}+\partial_{\nu}\partial_{\rho}A_{\mu}C_{\sigma}-\partial_{\mu}\partial_{\rho}A_{\nu}C_{\sigma}),\nonumber\\
\hat C_{\mu}&=&C_{\mu}+C^{\prime\prime}_{\mu}\nonumber\\
&=&C_{\mu}-\theta^{\rho\sigma}[A_{\rho}\partial_{\sigma}C_{\mu}-\partial_{\rho}A_{\mu}C_{\sigma}+\partial_{\mu}A_{\rho}C_{\sigma}],\nonumber\\
\hat\Lambda_{\mu}&=&\Lambda_{\mu}+\Lambda^{\prime\prime}_{\mu}\nonumber\\
&=&\Lambda_{\mu}-\theta^{\rho\sigma}[A_{\rho}\partial_{\sigma}\Lambda_{\mu}-\partial_{\rho}A_{\mu}\Lambda_{\sigma}+\partial_{\mu}A_{\rho}\Lambda_{\sigma}],
\label{sws}
\end{eqnarray}
Both solutions, however, are related by the field
redefinition
\begin{eqnarray}
B^{\prime\prime}_{\mu\nu}&=& B^{\prime}_{\mu\nu} +
\theta^{\alpha\beta} ( F_{\alpha [\mu} B_{\beta \nu]} - F_{\alpha
[\mu} \partial_\beta C_{\nu]} - C_\alpha
\partial_\beta F_{\mu\nu} ),\nonumber\\
C^{\prime\prime}_{~~\mu} &=&C^{\prime}_{~\mu} -
\theta^{\alpha\beta} C_\alpha F_{\beta\mu}.
\end{eqnarray}
Thus the SW map is unique up to field redefinitions. This happens
because more fields are available so that non-trivial field
redefinitions can be performed. The SW map for the primed fields
are the ones in Eqn.(\ref{sws1}).

By generalizing the gauge transformation of the 2-form theory
Eqns.(\ref {gtv}) and (\ref{gts}) to a p-form, with the C field a p-1 form
in d dimensions, coupled with the U(1) gauge field, the SW solution can
be generalized as
\begin{eqnarray}
{B}^{\prime}(B+\delta B,A+\delta A, C+\delta
C)-{B}^{\prime}(B,A,C)&=&d\Lambda^{\prime}+
\theta^{\rho\sigma}\partial_{\rho}A\partial_{\sigma}\Lambda-\theta^{\rho\sigma}\partial_{\rho}\lambda\partial_{\sigma}B\nonumber\\
{C}^{\prime}(C+\delta
C)-{C}^{\prime}(C)&=&\Lambda^{\prime}-\theta^{\rho\sigma}\partial_{\rho}\lambda\partial_{\sigma}C.
\label{swaji}
\end{eqnarray}
The solution for $B^{\prime}$, $C^{\prime}$ and
$\Lambda^{\prime}$ are,
\begin{eqnarray}
B^{\prime}&=&-\theta^{\rho\sigma}[A_{\rho}\partial_{\sigma}B+
dA_{\rho}\partial_{\sigma}C]+\theta^{\rho\sigma}\partial_{\rho}A\partial_{\sigma}C,
\nonumber\\
C^{\prime}&=&-\theta^{\rho\sigma}A_{\rho}\partial_{\sigma}C, \nonumber\\
\Lambda^{\prime}&=&-\theta^{\rho\sigma}A_{\rho}\partial_{\sigma}\Lambda.
\end{eqnarray}
Finally the p+1 form $\hat {\cal H}$ is,
\begin{equation}
\hat {\cal H}=-\theta^{\rho\sigma}dA_{\rho}\partial_{\sigma}{\cal
B}-\theta^{\rho\sigma}A_{\rho}\partial_{\sigma}d{\cal
B}+\theta^{\rho\sigma}\partial_{\rho}A\partial_{\sigma}{\cal B}.
\end{equation}}

The role of the C-field is interesting since it
 ensures the consistency of the KR equation, and also provides the KR
symmetry of action. Unlike the scalar and Maxwell's case the SW
mapped KR action does not have a structure which could be
interpreted as a gravitational coupling. This may mean that the
emergent gravity picture of the NC Maxwell theory may not be
viable for other gauge fields since the simplest
coupling to a Riemannian geometry was not obtained. This, however,
does not preclude the coupling to non-Riemannian geometries but
this is still under scrutiny. 

The dual scalar theory
obtained from SW mapped KR theory through Fradkin's method is
derived. The C field is present in it and interestingly enough its
field equation is  in accordance with the well known commutative
relation $\partial_{\mu}\phi\leftrightarrow {\tilde H_{\mu}}$
apart for O($\theta $) correction. The dual scalar theory also does not
have the structure of a gravity coupling, consistent with the
absence of a similar structure for the KR theory. Thus the duality
between the scalar and the tensor theory, present in the
commutative case, is not preserved in  the NC case. It is in agreement
with other studies on dualities where similar conclusions were
reached. The dispersion relation for the plane wave solution with
F constant was derived. An extension to massive 2-form gauge
theories, Stuckelberg and topologically massive theory is also
 an interesting subject. Work in these directions is in progress and
 will be reported later.

{\bf Acknowledgment} AKM and MS thank DST for support through a
project. The work of VOR is supported by CNPq, FAPESP and PROSUL
under contract 490134/2006-8.

\end{document}